\pdfoutput=1
\documentclass[a4paper]{jpconf}
\usepackage{graphicx,slashed}
\begin{document}
\title{New features of {\sc MadAnalysis}~5 for analysis design and reinterpretation}

\author{\underline{Eric Conte}$^1$, B\'eranger Dumont$^{2,3}$, Benjamin Fuks$^{4,5}$ and Thibaut Schmitt$^{5}$}
\address{
  $^{(1)}$ Groupe de Recherche de Physique des Hautes \'Energies (GRPHE),
       Universit\'e de Haute-Alsace, IUT Colmar, 34 rue du Grillenbreit BP 50568,
       68008 Colmar Cedex, France\\
  $^{(2)}$ Center for Theoretical Physics of the Universe, Institute for Basic Science (IBS), Daejeon 305-811, Korea\\
  $^{(3)}$ LPSC, Universit\'e Grenoble-Alpes, CNRS/IN2P3, 53 Avenue des Martyrs, F-38026 Grenoble, France\\
  $^{(4)}$ CERN, PH-TH, CH-1211 Geneva 23, Switzerland\\
  $^{(5)}$ Institut Pluridisciplinaire Hubert Curien/D\'epartement Recherches Subatomiques, 
    Universit\'e de Strasbourg/CNRS-IN2P3, 23 Rue du Loess, F-67037 Strasbourg, France}
\ead{eric.conte@iphc.cnrs.fr, beranger.dumont@lpsc.in2p3.fr, fuks@cern.ch, thibaut.schmitt@iphc.cnrs.fr}

\begin{abstract}
  We present \textsc{MadAnalysis 5}, an analysis package dedicated to
  phenomenological studies of simulated collisions occurring in
  high-energy physics experiments. Within this framework, users are invited,
  through a user-friendly \textsc{Python} interpreter, to implement physics
  analyses in a very simple manner. A \textsc{C++} code is then automatically
  generated, compiled and executed. Very recently, the expert mode of the
  program has been extended so that analyses with multiple signal/control
  regions can be handled. Additional observables have also been included,
  and an interface to several fast detector
  simulation packages has been developed, one of them being a tune of the \textsc{Delphes 3}
  software. As a result, a recasting of existing ATLAS and CMS analyses can
  be achieved straightforwardly.
\end{abstract}

\section{{\sc MadAnalysis 5} in a nutshell}
While both LHC experiments are currently pushing limits on particles beyond the
Standard Model to higher and higher scales, tremendous
progress has been made in the development of Monte Carlo event generators
and satellite programs, in particular with respect to precision predictions,
new physics implementations and event analysis. The public
\textsc{MadAnalysis}~5 package~\cite{Conte:2012fm,Conte:2014zja}
addresses this last aspect and provides a framework for analyzing events
generated either at the parton-level, after hadronization or after
detector simulation.
The program hence allows one to
efficiently design and recast LHC analyses. The user can in this way
investigate any physics model and determine the LHC sensitivity to
its signatures by either conceiving a novel analysis or recasting existing ATLAS
and CMS studies.
\textsc{MadAnalysis}~5 starts by reading event samples as generated by any Monte Carlo
event generator that satisfies community-endorsed output formats. Next, it
applies selection cuts and computes differential distributions as
requested by the user. In its normal mode of running,
the results are represented under the form of
histograms and cut-flow tables that are collected
within {\sc Html} and \LaTeX\ reports. In the expert mode of the program,
they are presented in text files compliant with the
{\sc Saf} format, an {\sc Xml}-inspired format internally used by
{\sc MadAnalysis}~5 that has been described in Ref.~\cite{Conte:2014zja}. Moreover,
events can be preprocessed. For instance, the application of a
jet-clustering algorithm or the
simulation of a detector response can be achieved effortlessly before the execution
of the analysis itself, and preprocessed events can be possibly saved in output files.

\section{Main concepts}
We recall in this section the main concepts of the \textsc{MadAnalysis 5} package.
We refer to Refs.~\cite{Conte:2012fm,Conte:2014zja} for more details.
\begin{itemize}
\item \textbf{Universal}:
  One can in the same way design new phenomenological
  analyses and recast existing LHC studies, and a given analysis can
  be equally applied on parton-level, hadron-level or reconstructed-level
  events stored under the \textsc{StdHep}~\cite{stdhep},
  {\sc HepMc}~\cite{Dobbs:2001ck}, {\sc Lhe}~\cite{Boos:2001cv,Alwall:2006yp},
  {\sc Lhco}~\cite{lhco} or {\sc Root}~\cite{Brun:1997pa} format.
\item \textbf{User friendly}: In the normal mode of running,
  the user designs her/his analysis by interacting with a
  {\sc Python} console through a metalanguage designed to be intuitive.
  The syntax is moreover easy to handle thanks to tab completion and in-line help.
\item \textbf{Efficient}: The {\sc Python} module of the program
  takes care of exporting the analysis encoded with the {\sc MadAnalysis}~5
  metalanguage to an optimized {\sc C++} program readily to be compiled and executed.
\item \textbf{Flexible}: \textsc{MadAnalysis 5} is shipped with a series
  of basic (transverse momentum, \textit{etc.})~and more sophisticated ($\alpha_T$, $M_{T2}$,
  {\it etc.})~built-in observables that can be used in analysis design. Operations on particle
  four-momenta are also available. For more complicated selections, the user can
  directly write his/her analysis in {\sc C++} (the so-called expert mode of the tool)
  and implement the desired observables.
\item \textbf{Interfacing}: \textsc{MadAnalysis 5} is interfaced to
  \textsc{Zlib}~\cite{ZLIB}, {\sc Root}~\cite{Brun:1997pa},
  \textsc{FastJet}~\cite{Cacciari:2011ma} (and \textsc{FastJet-contrib}) and
  \textsc{Delphes}~\cite{deFavereau:2013fsa}. The installation of these packages
  can be performed automatically from the {\sc Python} console. The program also contains
  the so-called \textsc{MA5tune} of {\sc Delphes}~\cite{Dumont:2014tja} (see Section~\ref{sec:fastsim}).
\end{itemize}

\section{Analysis design with the {\sc MadAnalysis}~5 metalanguage}
From the {\sc Python} console, the user is invited to describe its
selection and configure \textsc{MadAnalysis}~5 through intuitive and
human-understandable commands that inherit from the \textsc{Python} syntax.
As an exhaustive review of this metalanguage is beyond the scope of this
paper, we focus on a detailed illustrative example and refer to Ref.~\cite{Conte:2012fm} for more
information. We consider a simple analysis defined by the following commands:
\begin{verbatim}
  [0] ma5> import DrellYan*.hep as dy
  [1] ma5> import ttbar*.hepmc.gz as tt
  [2] ma5> set detector.fastsim.package = fastjet
  [3] ma5> set detector.fastsim.algo = kt
  [4] ma5> plot MET
  [5] ma5> define mu = mu+ mu-
  [6] ma5> select (mu) PT > 25
  [7] ma5> plot M(mu+ mu-)
  [8] ma5> set main.outputfile = output.lhco
  [9] ma5> submit
\end{verbatim}
The first two commands (lines 0 and 1) enable the import of the event samples to be considered.
We recall that \textsc{MadAnalysis 5} can read samples provided in
the \textsc{StdHep}, {\sc HepMc}, {\sc Lhe}, {\sc Lhco} or {\sc Root} format,
while compressed files can be used if the \textsc{Zlib} package is available.
The next two lines (lines 2 and 3) indicate how to use the interface to
the {\sc FastJet} program and select the $k_T$ algorithm for jet reconstruction
(see also Section~\ref{sec:fastsim}). The computation of two differential
distributions, namely the missing transverse energy and the muon-pair
invariant mass spectra, are then requested after enforcing the
selection of muons with a transverse momentum of at least 25~GeV (lines 4 to 7).
The next command (line 8) requires \textsc{MadAnalysis 5} to save the reconstructed events in
the {\sc Lhco} format, the {\sc Root} and {\sc Lhe} formats being also available.
The final command (line 9) requests the program to operate.
A {\sc C++} code corresponding to the analysis is generated and executed.
The results (histograms and selection efficiencies) are collected by a report generator based on
{\sc Root} (other graphical components will be available in the next release of
{\sc MadAnalysis}~5) and further exported into {\sc Html} and \LaTeX\ reports.

\section{Detector fast-simulation packages}
\label{sec:fastsim}
The \textsc{MadAnalysis 5} package offers the option to install, configure and launch
the fast simulation of a detector response on simulated events given as input in order to
mimic instrumental and experimental effects and increase the realism of the simulation.
Three possible choice are available, a very simple (but often sufficient for, \textit{e.g.},
generator-level based phenomenology) and fast simulation based on {\sc FastJet}, and
two more sophisticated and slower simulations based on {\sc Delphes}~3.
\begin{itemize}
  \item \textsc{MadAnalysis 5} is shipped with its own detector simulation based on \textsc{FastJet}.
  All the jet-clustering algorithms included in \textsc{FastJet} can be applied to hadron-level events,
  and the user can further switch on several detector effects like
  reconstruction efficiencies, resolution, misidentification rates, {\it etc.}
\item \textsc{MadAnalysis 5} is interfaced to \textsc{Delphes 3}, which offers a realistic description
  of the ATLAS and CMS detectors. Default configuration can be used, but the user is also
  invited to tune the {\sc Delphes} setup according to his/her needs. Pile-up effects can
  be included.
\item A modified version of \textsc{Delphes 3}, the so-called \textsc{Delphes-MA5tune}, can
  be employed. Based on {\sc Delphes}, it includes the
  calculation of new observables and optimizes the size of the output file. The main improvements
  address lepton and photon isolation. Unlike the official package, isolation variables
  are saved in the output file so that isolation requirements can be implemented at the analysis level
  instead of at the detector simulation level.
\end{itemize}

\section{Recasting existing LHC analyses with the {\sc MadAnalysis} 5 expert mode}
The \textit{expert mode} of the program offers the possibility to
write an analysis directly in {\sc C++}, using the \textsc{MadAnalysis}~5 framework.
It is developer-friendly, and, in addition, a large collection of
methods dedicated to common high-energy physics issues are available, as well as various
services allowing one, for instance, to produce cut-flow charts and/or histograms.
A detailed manual can be found in Ref.~\cite{Conte:2014zja}.
With version 1.1.10 onwards, one can also recast any cut-based ATLAS
and CMS analysis possibly containing multiple subanalyses or regions.
This allows reinterpretation of experimental results
under the approximation (compared to a real experimental setup)
of using a fast, simple, detector simulation software.
Our framework offers hence a way to work out the implications of the LHC
for any new physics model, derive
limits in a realistic-enough manner,
point out possible loopholes in the current searches and help to design
future analyses. It could also be used to improve the content and realism
of fast detector simulation packages.

The adopted strategy consists in first applying the \textsc{Delphes-MA5tune} fast detector
simulation on signal events stored under the {\sc StdHep} or {\sc HepMc} format. Secondly,
a reimplementation of a public ATLAS or CMS analysis is executed
to derive the number of selected signal events from the input samples.
A {\sc Python} script finally extracts limits by confronting the numbers associated with
the more sensitive signal region to the Standard Model expectation provided in the
experimental publication. In the case where a user would need a more
sophisticated procedure, he/she is welcome to extract directly the limits from
the {\sc MadAnalysis} output independently.
Since the validation of reimplemented analyses can
sometimes be complicated (see, \textit{e.g.}, Refs.~\cite{Dumont:2014tja,Kraml:2012sg}),
validated codes can be shared through our public analysis database~\cite{Dumont:2014tja}
and submitted to {\sc Inspire}~\cite{inspire}, that additionally assigns each submission a
DOI~\cite{doi}.

\section{Selection of results}
In order to demonstrate the performances of the program, we focus on two examples.
A first analysis is related to the search for a multileptonic signature of left-right symmetric
supersymmetry (LRSUSY) in simulated LHC collisions at a center-of-mass energy of
13~TeV~\cite{Alloul:2013fra}. A second example is devoted to the recast of the
CMS-SUS-13-012 analysis focusing on squark and gluino searches in multijet
events in collisions at a center-of-mass energy of 8~TeV~\cite{Chatrchyan:2014lfa,recastcmssus}.

\subsection{Prospective phenomenology in the left-right symmetric supersymmetric context}
\begin{figure}
\begin{minipage}[b]{16pc}
\includegraphics[width=16pc]{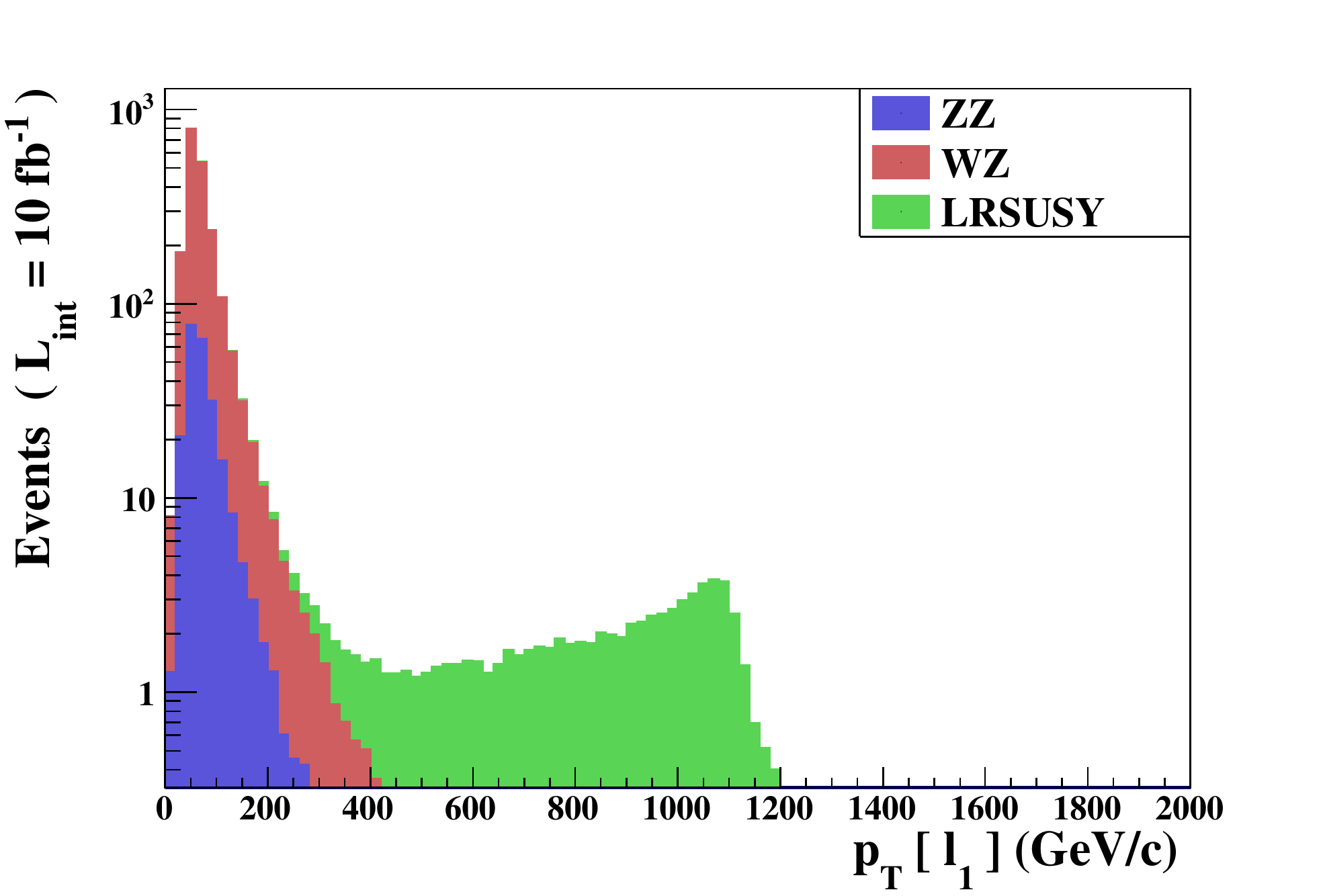}
\caption{\small \label{fig:plot1}Transverse-momentum distribution of the hardest lepton for the LRSUSY signal
  and the diboson background.}
\end{minipage}\hspace{2pc}%
\begin{minipage}[b]{20pc}
\includegraphics[width=20pc]{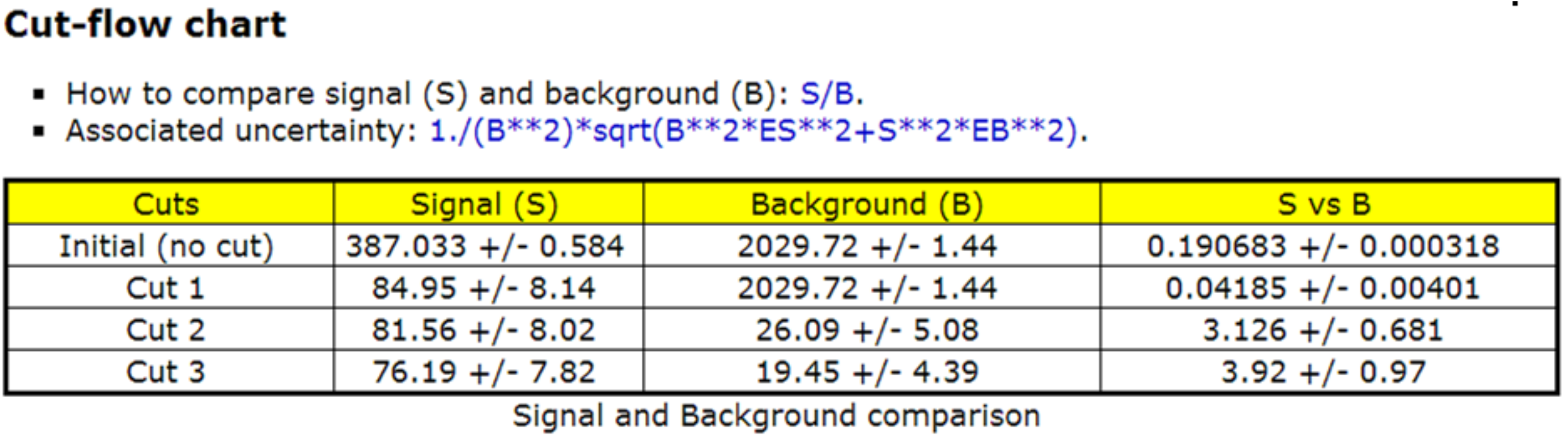} \vspace{0.2cm}
\caption{\small \label{fig:cutflow} Snippet of the \textsc{MadAnalysis 5} report with the cut-flow chart associated with the selection
described in the text. The figure of merit consists of the ratio of the number of signal to the number of background events.}
\end{minipage}
\end{figure}

LRSUSY predicts additional gaugino and higgsino states that can potentially lead to signals
in multileptonic events to be produced at the LHC. In the present example, we perform a simple selection
in order to extract the LRSUSY signal from the Standard Model $WZ$ and $ZZ$ background. In the
{\sc MadAnalysis}~5 metalanguage, the analysis reads
\begin{verbatim}
  ma5> define l = e+ e- mu+ mu-
  ma5> select N(l) >= 3
  ma5> plot PT(l[1]) 200 0 2000 [ logY ]
  ma5> select PT(l) > 200
  ma5> select MET > 50
\end{verbatim}
Events containing at least three charged leptons are selected. The
transverse momentum ($p_T$) distribution of the hardest lepton is then calculated,
the results being shown in Figure~\ref{fig:plot1}.
The signal exhibiting a harder spectrum, the $p_T$ of the hardest lepton
is constrained to be larger than 200~GeV, while at
least 50~GeV of missing transverse energy (assumed to be carried
by the lightest supersymmetric state that is stable and escapes detection)
is demanded. The corresponding cut-flow
table is presented on Figure~\ref{fig:cutflow}, that shows
that the selection yields a figure of merit of about $4\sigma$.

\subsection{Example of recast analysis}
\begin{figure}
\includegraphics[width=.47\columnwidth]{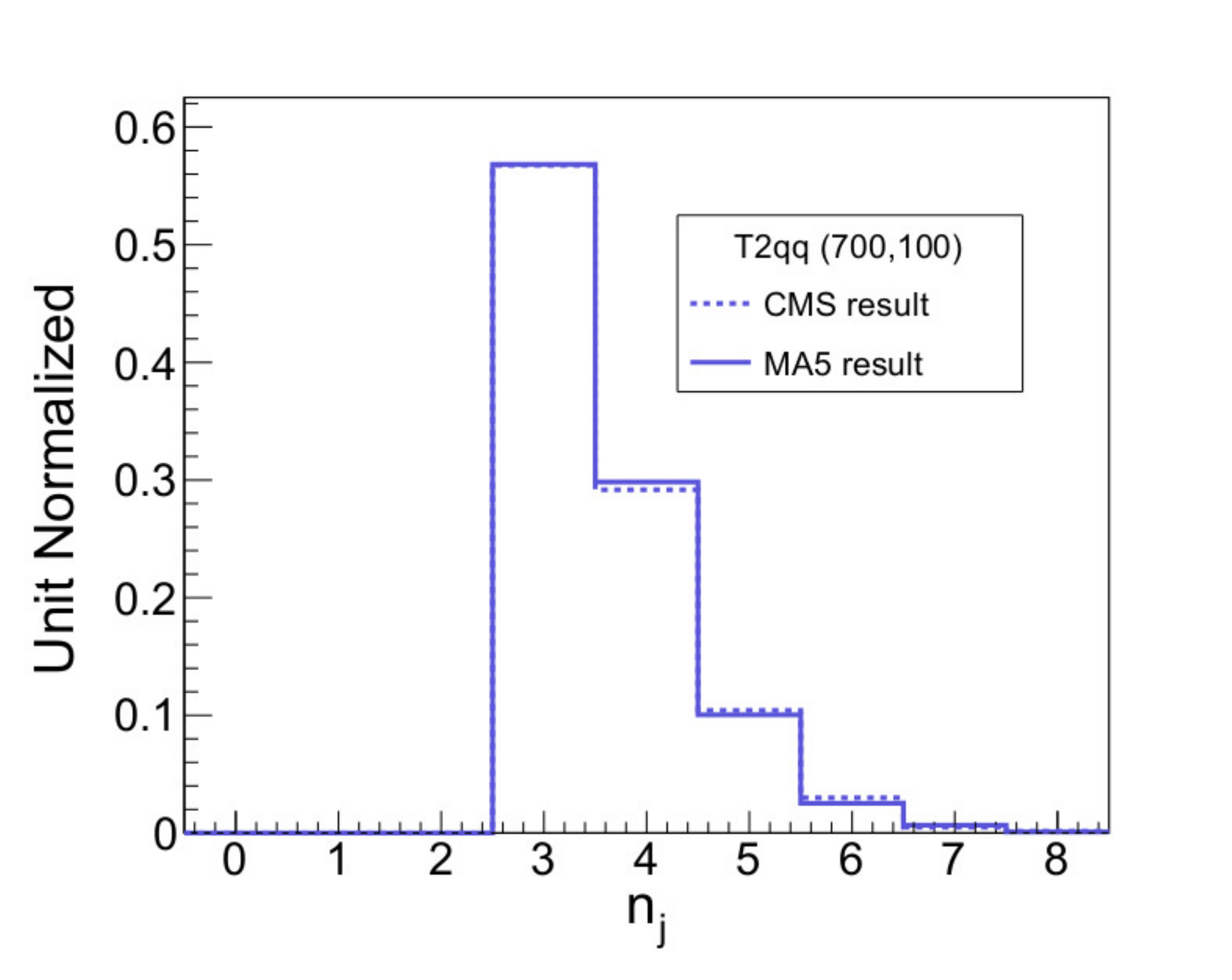}
\includegraphics[width=.47\columnwidth]{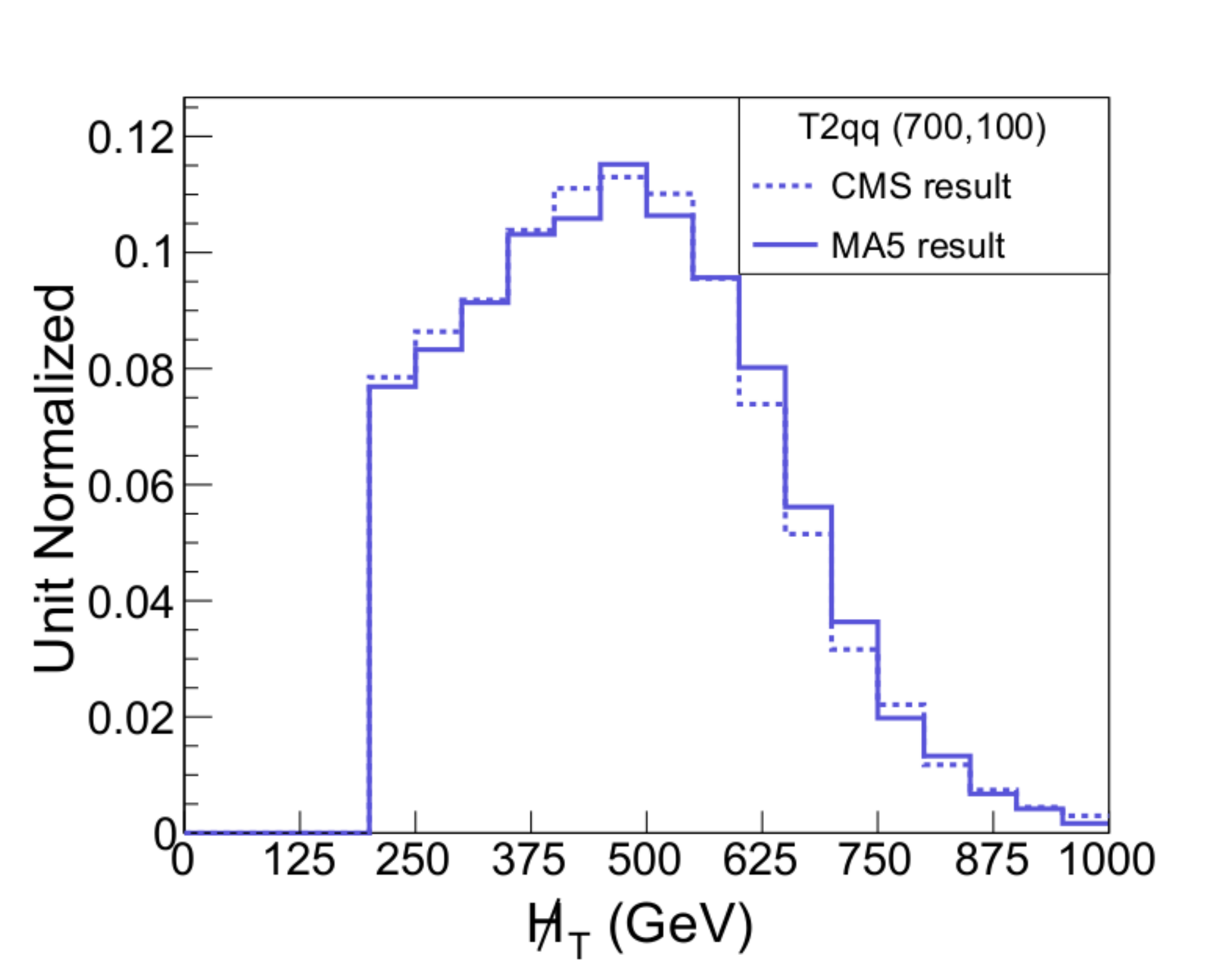}
\vspace*{-.2cm}
\caption{\small \label{fig:plot3}Comparison between CMS and MA5 for the distribution in
the jet multiplicity (left) and in the $\slashed{H}_T$ variable (right)
for a specific signal scenario and after the baseline selection described in Ref.~\cite{Chatrchyan:2014lfa}.}
\end{figure}

A recasting of the CMS-SUS-12-012 analysis~\cite{Chatrchyan:2014lfa} has been recently
performed, the validated {\sc MadAnalysis}~5 code being available in Ref.~\cite{recastcmssus}.
This validation has been achieved by comparing CMS and
{\sc MadAnalysis}~5 (MA5) results, both at the cut-flow level
and for various observables. Two examples are shown in Figure~\ref{fig:plot3},
where we present distributions for the jet multiplicity and the missing transverse hadronic energy
$\slashed{H}_T$ in the case of the \textsl{T2qq} simplified model~\cite{Chatrchyan:2013sza}. We
confront MA5 predictions to CMS official results and observe a good agreement.

\section{Summary}
The public package \textsc{MadAnalysis~5} offers a simple and efficient way
to design and recast LHC analyses. The user can either benefit from its
\textsc{Python} console and dedicated metalanguage to implement his/her
analysis, or make use of the expert mode of the program which provides a
flexible developer-friendly C++ environment. We have also discussed detector simulation
within the {\sc MadAnalysis}~5 framework, presenting the \textsc{Delphes-MA5tune}
simulation of a detector response, and introduced our recently initiated database
of reimplementations of LHC analyses.

\section*{Acknowledgments}
We thank the organizers for setting up this nice conference.
This work has been partially supported by the French ANR projects
12-JS05-002-01 BATS@LHC and 12-BS05-0006 DMAstroLHC, by
the Theory-LHC France initiative of the CNRS/IN2P3 and
by the \textit{Investissement d'avenir, Labex ENIGMASS}.

\section*{References}
\bibliographystyle{iopart-num}
\bibliography{ma5}
\end{document}